\documentclass[12pt]{article}

\def\com#1#2{\Big[#1,#2\Big ]}

\def\be{\begin{equation}}
\def\ee{\end{equation}}
\newcommand{\bea}{\begin{eqnarray}}
\newcommand{\eea}{\end{eqnarray}}

\def\fr{\frac}

\def\d{\delta}
\def\e{\epsilon}

\def\l{\lambda}

\def\s{\sigma}

\def\d{\delta}

\def\p{\partial}

\def\nn{\noindent}
\def\no{\nonumber}

\begin{document}
%\draft
%\tightenlines
%\preprint{IMSc/2002/11/39 }
%\smallskip\hbox{hep-th/}
\title{{\small\hfill IP/BBSR/2004-6}\\ 
Supersymmetric Quantum Mechanics on Non-Commutative Plane}
\author{E. Harikumar\thanks{hari@iopb.res.in}, 
V. Sunil Kumar\thanks{sunil@iopb.res.in} and 
Avinash Khare\thanks{khare@iopb.res.in}\\
Institute of Physics,\\Sachivalaya Marg, Bhubaneswar\\
Orissa, INDIA-751 005}
\maketitle
\begin{abstract}  

We study the Pauli equation on non-commutative plane. It is shown that
the Supersymmetry  algebra holds to all orders in the non-commutative parameter
$\theta$ in case the gyro-magnetic ratio $g$ is $2$.  Using
Seiberg-Witten map, the first order in $\theta$ correction to the
spectrum is obtained in the case of uniform magnetic field. We find
that the eigenstates in the non-commutative case are identical to the
commutative case provided the magnetic field $B$ is everywhere
replaced by $B(1+B\theta)$.

\end{abstract}
\medskip
\nn {PACS Numbers: 11.10.Nx, 02.40.Gh, 11.30.Pb}\\ 
\nn Keywords: non-commutative plane, Supersymmetry, Pauli equation\,\\
%%%%%%%%%%%%%%%%%%%%%%%%%%
\newpage
\section{Introduction}\label{intro}

Different aspects of the field theories formulated on non-commutative
spaces \cite{con} have been studied in the last few years. These
studies are motivated by the developments in string theories and also
from the attempts to understand the renormalization program \cite{sw}
. Studies in quantum gravity and matrix models \cite{mat} brought out
interesting aspects of non-commutative spaces.  Formulation of gauge
theories on non-commutative space shows many interesting features
\cite{gi} and have been widely investigated \cite{mdnn}. For example,
the problem of spin 1/2 particles in the constant electro-magnetic 
background has been analyzed \cite{gau,jab1} in the context of 
non-commutative QED. In particular, in \cite{gau} Schwinger's
world-line formulation has been worked out while in \cite{jab1} a 
direct solution of Dirac equation has been studied.It is interesting 
to note that in non-commutative QED, particle pair production is
unaltered at the tree level but have nontrivial effect at one loop
level\cite{jab1}. Further, as
shown in \cite{jab2}, this effect is enhanced if one considers
correction to $g-2$ in such a theory.

In the context of Non-locality
introduced by the non-commutativity of the space leads to the mixing
of UV and IR divergences in field theories formulated on
non-commutative spaces \cite{min}. The UV/IR mixing crucially affects
the renormalisability of the theory \cite{min,iya}. Recently
supersymmetric models on non-commutative spaces have been constructed
and the renormalisability of these model have been studied \cite{riv}.

Many interesting quantum mechanical problems on non-commutative (NC)
spaces have been investigated and the effect of non-commutativity on
the observables has been analyzed \cite{ncqm}. In particular, the
Landau problem has been formulated and studied extensively on
non-commutative settings \cite{ncqm}. But these results seems to differ
from one another and further, naively different gauge choices seen to
give different results. Recently using Seiberg-Witten map \cite{ sw,
sw1}, quantum Hall effect problem on non-commutative plane has been
studied \cite{bis}. Starting from the action for the Schr\"odinger
field coupled to $U(1)$ potential on non-commutative plane and using
Seiberg-Witten map (to the leading order in $\theta$), equivalent
commutative action for the case of uniform magnetic field was
obtained \cite{bis}.  The corresponding Schr\"odinger equation was then
derived and the gauge independence of the physical observables was
shown.

Recently renormalisability of non-commutative supersymmetric models
which do not involve gauge fields have been analyzed and it has been
argued that supersymmetric gauge theories on non-commutative spaces
have better renormalisability \cite{riv}. However so far as we are
aware of supersymmetric gauge theory models in NC settings have not
been investigated so far. The purpose of this paper is to take first
step in this direction. In this paper, we study the problem of a
charged particle moving in the non-commutative plane with
perpendicular magnetic field. Corresponding commutative problem, Pauli
equation, is known to have supersymmetry \cite{ak}.  We show that the
non-commutative generalization of Pauli equation has supersymmetry to
all orders in the non-commutative parameter $\theta$ and this
conclusion is valid irrespective of whether the magnetic field is
uniform or not. Using Seiberg-Witten map \cite{sw,sw1} we obtain first
order in $\theta$ correction to the spectrum in the case of uniform
magnetic field and explicitly demonstrate the gauge independence but
$\theta$ dependence of the spectrum. In section \ref{ncgauge} we
briefly present the essential features of non-commutative $U(1)$ gauge
theory and the Seiberg-Witten map. In section \ref{ncphe}, we
formulate the Pauli equation in non-commutative settings and show to
all orders in $\theta$ that the supersymmetry algebra is valid. The
exact spectrum valid to first order in $\theta$, is explicitly
obtained in Landau and symmetric gauges and it is shown that the
spectrum is the same as that in the commutative case provided the
uniform magnetic field $B$ is replaced everywhere by $B(1+B\theta)$.
We present concluding remarks in section \ref{con}.

\section{Non-commutative gauge theory}\label{ncgauge}

The non-commutative(NC) plane is defined by the co-ordinates ${\hat x}_i$ 
satisfying
\be
\com{{\hat x}_i}{{\hat x}_j}=i\e_{ij}\theta\,,~~i,j=1,2\,.
\label{com}
\ee
In the limit $\theta\rightarrow 0$, we recover commutative plane and
one requires the NC theory to reduce to usual commutative
theory in this limit. It is well known that the functions on NC 
space can be obtained using Moyal-Weyl correspondence. Corresponding to 
each function $f(x)$ on commutative space one associate a function
${\hat F}({\hat x})$ as
\be
{\hat F}({\hat x}) =\int dk~dx~ \exp^{ik({\hat x}-x)}~f(x)\,,
\ee
on NC space. On NC spaces the multiplication rule is modified 
and the new rule, the star product, is defined as
\be
{f}(x)*{ g}(x)=\exp^{\fr{i}{2}\theta^{ij}\partial_{i}^{x}\partial_{j}^{y}}
f(x)g(y)|_{x=y}\,.
\label{startp}
\ee

The gauge transformation of fields on NC space gets modified
because of this new multiplication rule and hence the structure of the gauge 
theory is much different from the commutative theory \cite{gi}. 
The non-commutative $U(1)$ theory is invariant under the $*$ gauge 
transformations
\bea
{\hat \Psi}(x)&\rightarrow&{\hat U}(x)*{\hat \Psi}(x)\no\\
{\hat A}_{i}&\rightarrow&{\hat U}(x)*{\hat A}_i*{\hat U}(x)^{\dagger}
+i{\hat U}(x)*\partial_i {\hat U}^{\dagger}\no\\\,,
\label{gt}
\eea
with ${\hat U}*{\hat U}^\dagger=1={\hat U}^\dagger *{\hat U}$ and the action 
of the covariant derivative is $D_i {\hat f}=\partial_i {\hat f} 
+i{\com{{\hat A}_i}{\hat f}}_*$.
The ${\hat U}(1)_*$ invariant field strength is given as
\be
{\hat F}_{ij}=\partial_i {\hat A}_j-\partial_j {\hat A}_i -i 
{\com{\hat{A}_i}{\hat{A}_j}}_*
\label{ncft}
\ee
and under $U(1)_*$ transformation this field strength transforms
 covariantly,i.e.,
\be
{\hat F}_{ij}\rightarrow {\hat U}*{\hat F}_{ij}*{\hat U}^\dagger\,.
\label{ftrans}
\ee
Unlike the commutative $U(1)$ case, in the NC $U(1)$ case the field
strength is in general not gauge invariant; only the integrated
expressions like action are gauge invariant. 
The Seiberg-Witten map \cite{sw} allows us to map the
gauge covariant objects like ${\hat F}_{ij}$ in terms of gauge
invariant objects in the commutative space. This map allows us to
express the NC gauge theory in terms of the commutative gauge fields 
but with the explicit $\theta$ dependence which
behaves as a background field. Thus after this mapping, the usual
calculational tools of commutative theory can be applied.

The Seiberg-Witten (SW) map is obtained by demanding that the ordinary gauge
potentials (living in the commutative space) which are connected by a
gauge transformation are mapped to NC potentials which are connected
by the corresponding NC gauge transformation. This requirement stated as
\be
{\hat A}(A) +\d_{\hat\l}{\hat A}(A) ={\hat A}(A+\p\l)\,,
\ee
gives the equation
\be
{\hat A}_{i}^\prime(A+\p\l)-{\hat A}_{i}(A)-\p_i\hat\l
=-\e^{kl}\theta\p_k\l\p_lA_i\,.
\label{sweq}
\ee
Solving this equation, to the order $\theta$ gives the well-known
SW map
\bea
{\hat A}_i&=&A_i-\fr{1}{2}\e^{kl}\theta A_k(\p_lA_i+F_{li})\,,\label{swmp}\\
\hat\l&=&\l+\fr{1}{2}\e^{kl}\theta \p_k\l A_l\,,
\label{swm1}
\eea
and for the matter field similar requirement results
\be
\hat \psi=\psi-\fr{1}{2}\e^{kl}\theta A_k\p_l\psi\,.
\ee
From Eqn. (\ref{swmp}) it follows
\be
{\hat F}_{ij}=F_{ij}+\e^{kl}\theta(F_{ik}F_{jl}-A_k\p_l F_{ij})\,.
\label{fh}
\ee
Under the $U(1)$ transformation of the commutative gauge potential
$A_i\rightarrow A_i+\p_i\l$, using Eqns. (\ref{swmp},\ref{swm1},
\ref{fh}) we get
\be
{\hat\d}_{\hat\l}{\hat F}_{ij}=-\e^{kl}\p_k\l\p_l F_{ij}\,,
\label{varF}
\ee
where $F_{ij}$ is the usual commutative Maxwell field strength,
thereby demonstrating the non invariance of NC field strength under
$U(1)_*$ transformation except in the special case of uniform field
strength. In this special case one can show that the exact relationship
between ${\hat B}$ and $B$ (valid to all finite orders in
$\theta$) is \cite{sw}  
\be
{\hat B}=\fr{B}{(1-B\theta)}
\label{conF}
\ee
where ${\hat B}={\hat F}_{12}$. Note that this relationship is
obviously not valid in case $B\theta =1$. We might add that there is
an alternative approach using covariant coordinates \cite{mad}, and
Seiberg-Witten map can also be derived in that formalism.

\section{Pauli Hamiltonian in $R_{\theta}^2$}\label{ncphe}

In this section we study the problem of a charged particle moving in
non-commutative plane ($R_{\theta}^2$) with magnetic field in the
perpendicular direction of the plane. The corresponding commutative
problem is known to have supersymmetry \cite{ak}.  We first
briefly sketch the commutative supersymmetric problem. The Hamiltonian
describing the system is given by ($\hbar=m=1$)
\be 
H=\fr{1}{2}\left[-(\p_i+iA_i)^2+\fr{g}{2}B\s_{3}\right]\,,
\label{ham}
\ee
where $i=1,2$ and $B=F_{12}$. This problem is known to have
supersymmetry (SUSY) in case the gyro-magnetic ratio, $g$, is $2$.
In particular, in this case the two hermitian supercharges are
\bea
Q_1&=&\fr{1}{\sqrt{2}}\left[i(\p_y+iA_y)\s_1-i(\p_x +iA_x)\s_2\right],\no\\
Q_2&=&\fr{1}{\sqrt{2}}\left[-i((\p_x +iA_x)\s_1
-i(\p_y+iA_y)\s_2\right]\,.
\eea
where $\s_i$ are Pauli matrices. In terms of these, the complex supercharge 
is defined to be 
\be
Q=-\fr{i}{2} (Q_1-iQ_2)\,.
\label{comq}
\ee
It is straight forward to see that the SUSY algebra 
\be
Q^2=0\,,~~~\left\{Q,Q^\dagger\right\}=H\,,~~\com{H}{Q}=0=\com{H}{Q^\dagger}\,,
\ee
is satisfied where $H$ is given in Eqn. (\ref{ham}) with $g$ being $2$. 

In this section we study generalization of this problem to NC
plane. In the NC generalization of quantum mechanical problems as well
as that of field theories it is known that one can either work
directly with NC variables or work with usual commutative variables
but with the usual products modified to star product
\cite{ncqm,jab}. The 2 dimensional central field problem, Landau
problem etc have been analyzed using the second approach recently. In
these papers the effect of star product has been absorbed in a
momentum dependent shift of the co-ordinates.  In symmetric gauge, the
Landau problem has been shown to be equivalent to NC central field
problem using this method \cite{ncqm}. But naively this approach seems
to give gauge dependent answers. In \cite{bis} the NC Hall effect
problem has been studied using the first approach and by employing the
SW map the NC action was re-expressed in terms of the commutative
variables and $\theta$ and the effect of non-commutativity was
studied. We feel this approach is better suited to study problems
involving gauge potential and we adopt this method in this paper.

We start with the Hamiltonian expressed in terms of the NC variables
\bea
{\hat H}&=&\fr{1}{2}\left[-(\p_i+i{\hat A}_i)^2+\fr{g}{2}{\hat B}_{3}\s_{3}\right]\no\\
&=& \left (
\begin{array}{cc} 
{\hat H}_2~&0\cr
0&{\hat H}_1 \cr
\end{array} \right)
\label{nch}
\eea
where hatted variables live on the NC plane and 
${\hat B}=\fr{1}{2}\e_{ij}{\hat F}_{ij}$ with ${\hat F}_{ij}$ given in
Eqn. (\ref{ncft}). The NC supercharges are 
\bea
{\hat Q}_1= \fr{1}{\sqrt{2}}\left (
\begin{array}{cc} 
0~&~-(\p_x+i{\hat A}_x)+i(\p_y+i{\hat A}_y)\cr
(\p_x+i{\hat A}_x)+i(\p_y+i{\hat A}_y)\cr&0 \cr
\end{array} \right)\no\\
{\hat Q}_2= \fr{1}{\sqrt{2}}\left (
\begin{array}{cc} 
0~&~-i(\p_x+i{\hat A}_x)-(\p_y+i{\hat A}_y)\cr
-i(\p_x+i{\hat A}_x)+(\p_y+i{\hat A}_y)\cr&0 \cr
\end{array} \right)~
\eea
From this we get the complex super charge (Eqn. \ref{comq}) to be
\bea
{\hat Q}&=& \left (
\begin{array}{cc} 
0~&~{\cal A}\cr
0&0 \cr
\end{array} \right)\no\\
{\rm~where~~~}
{\cal A}&=&\fr{1}{\sqrt{2}}\left[i(\p_x+i{\hat A}_x)+(\p_y +i{\hat A}_y)\right].
\label{cala}
\eea
It is straight forward to see that the SUSY algebra 
\be
\left\{{\hat Q},{\hat Q}\right\}_{*}=0,~~\left\{{\hat Q},{\hat
Q}^\dagger\right\}_{*} ={\hat H},~
\com{{\hat H}}{\hat Q}_{*} = 0 =\com{{\hat H}}{{\hat Q}^\dagger}_{*}
\label{susyalg}
\ee
is satisfied with $H$ as given in Eqn. (\ref{nch}) and $g=2$. Note that
this algebra is valid irrespective of whether the field strength is
uniform or not and also irrespective of the value of the NC parameter
$\theta$. Thus we have shown that in the NC settings, the Pauli
equation can be casted in the supersymmetric form provided $g$ is again $2$. 
This conclusion is valid to all orders in $\theta$.

Using Eqn. (\ref{cala}) and (\ref{susyalg}) the SUSY partner Hamiltonians can 
be expressed as
\be
{\hat H}_1={\cal A}^\dagger{\cal A}\,,~~~~~
{\hat H}_2={\cal A}{\cal A}^\dagger\,,
\ee
and the ground state wave function of ${\hat H}_1$ obeys
\be
{\cal A}\Psi_{0}^{1}=0\,,
\label{nca}
\ee
where ${\cal A}$ is given by Eq. (\ref{susyalg}).
 
We shall now specialize to the case of uniform magnetic field and solve
the NC Pauli equation in two different gauges. We obtain the spectrum
valid to first order in $\theta$ explicitly in both the Landau and
the symmetric gauges.
In the commutative case, when  the magnetic field is uniform, then in the
Landau gauge the corresponding gauge potentials are given by
\be
A_x=-By\,,~~A_y=0\,,
\label{nclan}
\ee
while in the symmetric gauge, they are given by
\be
A_x=-\fr{1}{2}By\,,~~A_y=\fr{1}{2}Bx\label{sym}\,.
\ee
These two gauges are connected by $U(1)$ gauge transformation
$A_i\rightarrow A_i+\p_i\l$ where the transformation parameter
$\lambda$ is given by 
\be
\l=-\fr{1}{2}Bxy\,.
\label{lam}
\ee

The NC potential in a given gauge can be expressed in terms of the
corresponding commutative
potential and $\theta$. For example, in the Landau gauge, to order
$\theta$, such a connection is easily obtained by using
Eqn.(\ref{swmp}) and one finds that 
\be
{\hat A}_{x}=-B(1+B\theta)y\,, ~~~~{\hat A}_{y}=0\,.
\label{swal}
\ee
On the other hand, using Eqns.(\ref{sym},\ref{lam},\ref{swal} and \ref{swm1}) 
in Eqn.(\ref{sweq}),
the NC gauge potential in the symmetric gauge is given by 
\be
{\hat A}_{x}=-\fr{1}{2}B(1+B\theta)y\,,~~~~~
{\hat A}_{y}=\fr{1}{2}B(1+B\theta)x\,.
\label{ncsym}
\ee
Using Eqns. (\ref{fh}),(\ref{varF}) and (\ref{conF}) it
is clear that corresponding to the commutative magnetic field $B$,
the NC magnetic field ${\hat B}$, to order $\theta$  is given by
\be
{\hat B}=B(1+B\theta)\,.
\label{bnc}
\ee

Using Eqs. (\ref{cala}), (\ref{nca}) and (\ref{swal}) one can now
immediately obtain the ground state wave function of $\tilde{H}$ in
the Landau gauge. We
find that
\be\label{gpsi}
\psi_{0}^{1}= \exp(ikx) \exp[-\fr{\rho^2}{2B(1+\theta B)}]\,
\ee
where $\rho=(By-k)$ and $k$ being any real number.  Similarly, using
the SW map, in the Landau gauge, the NC Pauli equation ${\hat H} \psi
= E\psi$ takes the form 
\be
\left[-\p_{y}^2-(\p_x-i(1+B\theta)By)^2+ B(1+B\theta)\s\right]\Psi
     =2 E\Psi\,,
\ee
where we have used Eqns. (\ref{swal}) and (\ref{bnc}). Thus
in this gauge, effectively the problem has become an one
dimensional one. With the redefinition $B^\prime=B(1+B\theta)$, we get
the same equation as in the commutative space \cite{ak} and hence  
the spectrum is given by
\be
E_n=(n+\fr{1}{2}+\fr{1}{2}\s)B(1+B\theta)\,,~~\s=\pm 1\,.
\label{spec1}
\ee 
Thus the energy eigenvalues, spectral spacing and density of states 
get a $\theta$ dependent modification. Further, 
as in the commutative case, the spectrum is again independent
of $k$ (see Eqn. (\ref{gpsi})). Thus the energy levels are continuously 
degenerate. 
%and not affected by the $\theta$ parameter. 

In the symmetric gauge, using Eqns. (\ref{ncsym}) and (\ref{bnc}),
the SW mapped Pauli equation becomes
\be
\left[
-(\p_x-\fr{i}{2}B^\prime y)^2-(\p_{y}+i\fr{i}{2}B^\prime x)^2
+ B^\prime \s\right]\Psi
     =2E\Psi\,.
\ee
where $B^\prime=B(1+B\theta).$ As in the previous case, 
the above equation has exactly the same form as in the commutative
symmetric gauge and hence the spectrum is
\be
E_n=(n+\fr{1}{2}+\fr{1}{2}\s+m+|m|)B(1+B\theta)\,,~~~\s=\pm 1\,,~~~m=0,\pm 1,\pm 2,..~.
\label{spec2}
\ee
Here also the energy eigenvalues, the spectral spacing as well the
density of states are modified due to the NC parameter $\theta$ while
the degeneracy is exactly the same as in the commutative case. In
particular, we see that all the energy levels including the ground
state are infinite-fold degenerate.

Thus to the first order in $\theta$, the spectrum while gauge
invariant, is affected by the non-commutativity and the NC parameter
$\theta$ appears explicitly in the energy eigenvalue expression.  We
see from Eqns. (\ref{spec1}) and (\ref{spec2}) that the ground state
energy for ${\hat H}_1$ is zero in both gauges and thus to first order
in $\theta$, SUSY continues to remain unbroken in the NC plane.

Before finishing this note, it is worth pointing out that in case we
assume that even in the NC case, in the Landau gauge, ${\hat A}_{y}=0$,
then using the exact SW map and using the exact relationship between 
$\hat{B}$ and $B$ as given by Eq.(\ref{conF}), it is easy to show that the
exact relationship between $\hat{A}_{x}$ and $A_x$ in the Landau gauge
is given by
\be\label{exact}
\hat{A}_{x} =\fr{A_x}{(1-B\theta)}= -\frac{By}{(1-B\theta)}\,,~~\hat{A}_{y}=0\,.
\ee
Hence it is now easy to show that to all orders in $\theta$, the
exact spectrum is given by 
\be\label{exasp}
E_n = (n+\frac{1}{2}+\frac{1}{2}\sigma){\hat B}\,,~~ \sigma = \pm 1\,,
\ee
where ${\hat B}$ is as given in Eqn.(\ref{conF}). The SUSY remains unbroken
to all orders in $\theta$.

\section{Conclusion}\label{con}

In this paper we have formulated and studied the non-commutative
generalization of the Pauli equation and have shown how even that can
be put in a supersymmetric setting.  We have shown to all orders in
$\theta$ that the SUSY algebra continues to hold in the NC case for
$g=2$.  Using SW map, we have obtained the equivalent Hamiltonian, to
first order in $\theta$, in terms of commutative variables and NC
parameter $\theta$. Using this SW mapped Hamiltonian, we have obtained
the energy eigenvalues in two different gauge choices in case the
magnetic field is uniform. The spectrum gets affected by the
non-commutativity and to the first order in $\theta$ we have shown
that all the energy eigenvalues are scaled by the factor of
$1+B\theta$. We have also shown that, to order $\theta$, the SUSY
remains unbroken in the NC plane. The spectrum is infinitely
degenerate as in the commutative plane and interestingly the
degeneracy does not depend on $\theta$ though the spectral spacing
does. Finally, assuming that even in the NC Landau gauge ${\hat
A}_y=0$, we have obtained the spectrum which is valid to all orders in
$\theta$.

We have used the SW map to re-express to order $\theta$, the NC Hamiltonian in terms of
commutative potential $A_i$ and $\theta$. 
It has been claimed \cite{sf} that in the the SW map the $\theta^2$
and higher order
terms are not unique. Since the SUSY is satisfied to all
orders in $\theta$, demanding the SUSY algebra
order by order in $\theta$
may allow us to single out the preferred choice of the higher order
$\theta$ terms in SW map.\\

\nn {\bf Acknowledgements}: One of us (EH) thank M. Sivakumar for useful 
discussions.


\begin{thebibliography}{99}
\bibitem{con} A. Connes, Noncommutative geometry, Academic Press, London, (1994).
\bibitem{sw} N. Seiberg and E. Witten, JHEP {\bf 09} (1999) 032.
\bibitem{mat}A. Connes, M. R. Douglas and A. Schwarz, 
JHEP {\bf 9802} (1998) 033.
\bibitem{gi} J. A. Harvey, `Topology of the Gauge Group in Noncommutative 
Gauge Theory', hep-th/0105242; C. Sochichiu, 
`Gauge Invariance and Noncommutativity', hep-th/0202014.
\bibitem{mdnn} M. R. Douglas and N. A. Nekrasov, Rev. Mod. Phys. {\bf 73} (2001) 977.
\bibitem{gau}L. Alvarez-Gaume and J.L.F. Barbon Int. J. Mod. Phys. 
{\bf A 16} (2001) 1123, hep-th/0006209.
\bibitem{jab1} N. Chair and M. M.  Sheikh-Jabbari,  Phys. Lett. {\bf B504} 
(2001) 141, hep-th/0009037.
\bibitem{jab2}I.F. Riad and M. M. Sheikh-Jabbari, JHEP {\bf 0008} (2000) 045, 
0008132.
\bibitem{min} S. Minwalla, M. V. Raamsdonk, N. Seiberg, JHEP {\bf 0002} 
(2000) 020.
\bibitem{iya}I. Y. Aref\'eva, D. M. Belov and A. S. Koshelev,
Phys. Lett. {\bf B476} (2000) 431; S. Sarkar, B. Sathiapalan,
JHEP {\bf 0105} (2001) 049, S. Sarkar, JHEP {\bf 0206} (2002) 003.
\bibitem{riv} H. O. Girotti, M. Gomes, V. O. Rivelles and A. J. da Silva,
Nucl. Phys. {\bf B587} (2000) 299; ibid, Int. J. Mod. Phys. {\bf A 17} (2002) 
1503.
\bibitem{ncqm}M. Chaichian, M. M. Sheikh-Jabbari and A. Tureanu,
Phys. Rev. Lett. {\bf 86} (2001) 2716; J. Gamboa, M. Loewe, F. Mendez and 
J. C. Rojas, Mod. Phys. Lett. {\bf A16} (2001) 2075; ibid, Int. J. Mod. Phys.
{\bf A 17} (2002) 2555;
S. Bellucci and A. Nersessian, Phys. Lett. {\bf B542} (2002) 
295-300. P. A. Horvathy, Annals Phys. {\bf 299} 
(2002) 128; H. R. Christiansen and F.A. Schaposnik, Phys. Rev. {\bf D65} 
(2002) 086005; V. P. Nair and A. P. Polychronakos, Phys. Lett. 
{\bf B505} (2001) 267; L. Mezincescu, `Star Operation in Quantum Mechanics',
hep-th/0007046.
\bibitem{sw1} A. A. Bichl, J. M. Grimstrup, L. Popp, M. Schweda and R. Wulkenhaar, `Deformed QED via Seiberg-Witten Map',hep-th/0102103
\bibitem{bis} B. Chakraborty, S. Gangopadhyay and A.Saha, 
`Quantum Hall effect on non-commutative plane through Seiberg-Witten map', 
hep-th/0312292.
\bibitem{ak} M. de Crombrugghe and V. Rittenberg, Ann. Phys. {\bf 151} (1983) 
99; A. Khare and J. Maharana, Nucl. Phys. {\bf B244} (1984) 409;
F. Cooper, A. Khare and U. Sukhatme, `Supersymmetry in Quantum Mechanics', (World Scientific, 2001).
\bibitem{mad}J. Madore, S. Schraml, P. Schupp and J. Wess, Eur. Phys. J 
{\bf C16} (2000) 161; R. Jackiw, S.-Y. Pi, Phys. Rev. Lett. {\bf 88} (2002) 
111603.
\bibitem{jab} A. Micu and M. M. Sheikh-Jabbari, JHEP {\bf 0101} (2001) 025.
\bibitem{sf}S. Fidanza, JHEP {\bf 0206} (2002) 016.
\end{thebibliography}
\end{document}